\documentclass[aps,prl,twocolumn,superscriptaddress]{revtex4}
\usepackage{epsf}
\begin{document}

\title{Nonlinear evolution of surface morphology in InAs/AlAs superlattices via surface diffusion}

\author{O. Caha}
\email{caha@physics.muni.cz}
\affiliation{Institute of Condensed
Matter Physics, Masaryk University, Kotl\'a\v rsk\'a 2, 61137
Brno, Czech Republic}
\author{V. Hol\'{y}}
\affiliation{Department of Electronic Structures, Charles
University, Praha, Czech Republic}
\author{Kevin E. Bassler}
\affiliation{Department of Physics, University of Houston, Houston, Texas 77204-5005, USA}

\date{\today}

\begin{abstract}
Continuum simulations of self-organized lateral compositional
modulation growth in InAs/AlAs short-period superlattices on InP
substrate are presented. Results of the simulations correspond
quantitatively to the results of synchrotron x-ray diffraction
experiments. The time evolution of the compositional modulation
during epitaxial growth can be explained only including a
nonlinear dependence of the elastic energy of the growing
epitaxial layer on its thickness. From the fit of the experimental
data to the growth simulations we have determined the parameters
of this nonlinear dependence. It was found that the modulation amplitude don't depend on the values of the surface diffusion
constants of particular elements.
\end{abstract}

\pacs{68.65.Ac;81.16.Dn;61.10.Nz}
\maketitle

The process of self-organization during the growth of
semiconductor epitaxial nanostructures is described using two
different models \cite{review}. If there is a high density of
monolayer steps on the vicinal surface (the crystallographic
miscut angle is larger than approx. 1$^\circ$), a step-bunching
instability occurs \cite{bai}, but if the density of the the monolayer
steps is low, a self-organized growth of two-dimensional or
three-dimensional islands takes place. The latter process occurs,
if the reduction of the strain energy due to an elastic relaxation
of internal stresses in the islands outweighs the corresponding
increase of the surface energy (morphological
Asaro-Tiller-Grinfeld (ATG) instability
\cite{asaro,grinfeld,srolovitz}).

The cited papers analyzed the self-organization process in a
linearized approach from which a critical wavelength of the
surface corrugation follows as function of material parameters.
The exact nonlinear equation of the surface evolution was studied
by Yang and Srolovitz \cite{yang} and Spencer and Meiron
\cite{spencer1} for the case of a semi-infinite substrate. It was
found that the shape of an initially harmonic surface waviness
changes and a sequence of deep cusps is created.
This behavior was observed using scanning electron microscopy
(see, e.g., \cite{jesson1994}).

The physical properties of very thin layers (down to few
monolayers) differ from the properties of the bulk. This
difference leads to the creation of a stable two-dimensional layer
at the surface (wetting layer) in the first stage of the
Stranski-Krastanov growth mode. The occurrence of this  so-called
"wetting-effect" can be explained by a nonlinear dependence of the
elastic energy density on the layer thickness
\cite{spencer1999,tersoff1991}. Simulations showed that the
wetting-effect suppresses the growth of the cusps and subsequently
it leads to the formation of surface islands \cite{savina}. These islands are
unstable and coalesce (the Ostwald ripening \cite{ostwald})
\cite{seifert1996}. However, numerical growth simulations indicate
that an anisotropy of the surface energy limits the
ripening process and causes the creation of a nearly
homogeneous array of islands (see Ref. \cite{eisenberg2}, among
others, and the citations therein).

The evolution of the surface morphology of multilayers has been
studied only in a linearized approach so far
\cite{shilkrot,huang}. From this approach, an unlimited growth of
the modulation amplitude follows, which does not correspond to the
experimentally observed stabilization of the modulation amplitude
during the growth. The aim of this paper is to describe this
stabilizing effect using the exact nonlinear equation of growth
including the wetting-effect. We have simulated the time evolution
of the spontaneous lateral modulation of layer thicknesses in
short-period semiconductor superlattices and we have found that
the evolution of the modulation amplitude
quantitatively corresponds to the results obtained by x-ray
scattering measurements.

We have studied a series of InAs/AlAs superlattices grown by
molecular beam epitaxy on InP (001) substrates. The substrate was
covered by a 100\, nm thick In$_x$Al$_{1-x}$As buffer layer with
the same chemical composition as the average composition of the
InAs/AlAs stack. The nominal thicknesses of both InAs and AlAs
layers were 1.9 monolayers (mL) in all samples. The samples were
prepared in a series with 2, 5, 10, and 20 superlattice periods.
The growth temperature was 530$^\circ$C and the growth rate
0.5~mL/s. The details of the growth can be found elsewhere
\cite{ahrenkiel}.

\begin{figure}
 \epsfxsize=70mm
 \epsfbox{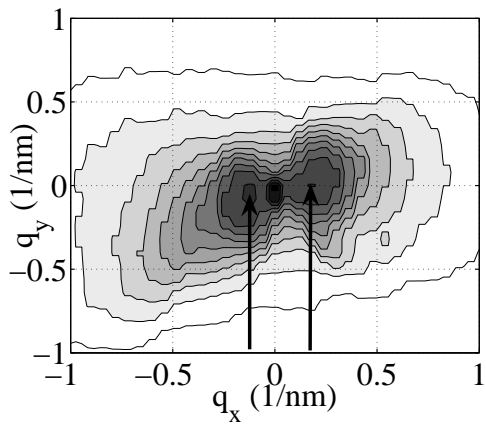}
 \caption{Diffusely scattered intensity measured around the 400 reciprocal lattice point
  on the sample with 20 superlattice periods. The contour step is 10$^0.25$.
  The diffraction vector is along the $q_x$ axis. The arrows show lateral intensity maxima mentioned in the text.
  }
 \label{f1}
\end{figure}

X-ray grazing-incidence diffraction (GID) measurements were
carried out at beamline ID01 at European Synchrotron Radiation
Facility (ESRF) in Grenoble using the x-ray wavelength
1.54\,\AA{}. We have measured the diffusely scattered intensity
distributions around 400 and 040 reciprocal lattice points in the
$q_xq_y$ plane parallel to the sample surface. Two first-order
lateral satellite maxima were observed at samples with 5 and more
superlattice periods. An example of this experimental intensity
map is plotted in Fig. \ref{f1}.
The distance of
the lateral satellites from the specular crystal-truncation rod at
$q_{x,y}=0$ is inversely proportional to the lateral modulation
period; the period remains constant for all samples and was
determined to be $(267\pm15)$\,\AA{}. From the position of the
satellite maxima in the $q_xq_y$ plane we determined the
modulation directions close to [310] and [$\bar{1}$30]. The
amplitudes of the satellites increase with the growing number of
the superlattice periods, while their widths decrease. This
indicates that during the epitaxial growth the periodicity of the
lateral modulation improves and its amplitude increases. The
details of the experimental setup and results are described in the
previous paper \cite{nasclanek}, where we have determined the
time-dependence of the modulation amplitude from the x-ray data.

\begin{figure}
 \epsfxsize=70mm
 \epsfbox{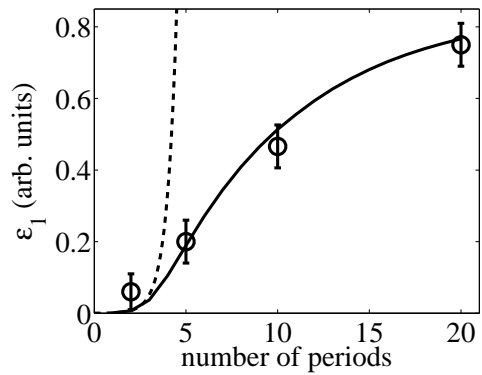}
 \caption{The dependence of
 $\varepsilon_1$ on the number of superlattice periods; this dependence
 describes the time evolution of the modulation during the growth.
 The circles with error bars are the experimental points obtained from the x-ray data,
 the full line represents the simulations.
 The dashed line is the evolution of $\varepsilon_1$ calculated in the linearized case \cite{shilkrot,huang}.
  }
 \label{f3}
\end{figure}

The diffuse scattered intensity can be calculated using the
formula
\begin{equation}
I_{\rm diff}({\bf q})=A\left|\int{\rm d}^3{\bf r} \chi_{\bf
h}({\bf r}){\rm e}^{-i{\bf h}\cdot{\bf u}({\bf r})}\right|^2,
\end{equation}
where $A$ is a constant, $\chi_{\bf h}$ is the crystal
polarizability, ${\bf h}$ is the diffraction vector, and ${\bf u}$
is the displacement vector. Using the procedure described in the
previous work \cite{nasclanek}, we have extracted the correlation
function $\varepsilon({\bf x}-{\bf x}')=\langle
(c({x})-c_0)(c({x}')-c_0)\rangle,$ from the measured data, where
$c({x})$ is the local InAs concentration averaged along the growth
direction $z$ and $c_0$ is the average InAs concentration. The
dependence of the first coefficient $\varepsilon_1$ of the Fourier series of
$\varepsilon$ obtained from the experimental data is plotted
in Fig. \ref{f3}; this coefficient corresponds to the modulation
amplitude.

The evolution of the surface described by the function $z=h(x,t)$
is driven by the surface diffusion and it can be described by the
equation \cite{mullins}
\begin{equation}
\frac{\partial h(x,t)}{\partial t}=\frac{D_s\theta\Omega}{k_BT}
\nabla^2\mu(x,t)+F+\eta(x,t),
\end{equation}
where $D_s$ the surface diffusion coefficient, $\theta$ is number
of atoms per unit area on the surface, $\Omega$ is atomic volume,
$k_B$ is the Boltzmann
constant, $T$ is the temperature, $\mu$ is the chemical potential on the surface, $F$ is the deposition rate, and
$\eta$ is the deposition and diffusion random noise. The model is
assumed to be independent on the third coordinate $y$ and hence
only one-dimensional surface modulation in a two-dimensional
$(x,z)$ space can be simulated. This approach is well justified
for quantum wires or for strongly elongated quantum dots. For the
quantum dots of a more symmetric shape this simulation can give
limited information only. This effect is discussed below.

The chemical potential $\mu$ can be expressed as
\cite{eisenberg}
\begin{equation}
\mu(x,t)=\mu_0+\gamma\kappa+\left.\frac{1}{2}C_{jklm}
\epsilon_{jk}\epsilon_{lm}\right|_{z=h(x)}+\frac{{\rm
d}f_{el}^{(0)}(h)}{{\rm d}h},
\end{equation}
where $\mu_0$ is the chemical potential of an ideally flat
unstrained surface, $\gamma$ is the surface tension, $\kappa$ is
the surface curvature, $C_{jklm}$ are the components of the
elastic tensor of the material, and $\epsilon_{jk}$ is the
strain tensor. The function $f_{el}^{(0)}(h)$ describes a
dependence of the elastic energy density on the layer thickness
$h$ giving rise to a wetting-effect. For a Ge layer on Si, this
function was approximated by the exponential function
\cite{eisenberg}
\begin{equation}
f_{el}^{(0)}(h)\approx 0.05\times E_S(1-\exp(-h/h_{\rm ml})),
\end{equation}
where $E_S$ is strain energy density in thick flat layer and
$h_{\rm ml}$ is thickness of one monolayer. We have used an
analogous formula
\begin{equation}
f_{el}^{(0)}(h)= E_W(1-\exp(-h/h_W)),
\label{wett}
\end{equation}
where $E_W$ and $h_W$ are parameters depending on the lattice
misfit and elastic constants of the layer.

The strain energy was calculated by a direct solution of the
linear elasticity equations using the boundary integral method.
The method used is a multiple layer extension of the method in Ref. \cite{yang}
for an isotropic continuum with periodic boundary conditions. The
boundary conditions on the internal interfaces are described in
Ref. \cite{luo2}, the nearly singular integrals were
calculated using method developed in Ref. \cite{luo}.

\begin{figure}
 \epsfxsize=70mm
 \epsfbox{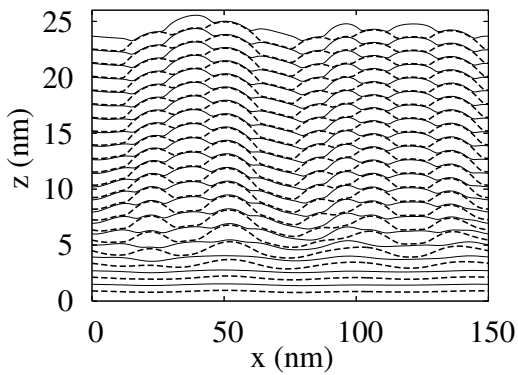}
 \caption{Simulated interface profiles in a InAs/AlAs superlattice. Thick dashed lines
 denote the interfaces between InAs (under) and AlAs (above).
 Thin solid lines denote the interfaces between AlAs and InAs.
  }
 \label{f2}
\end{figure}

The simulations have been performed with known material parameters.
The surface energy $\gamma$, calculated from the first principles, 
was taken from the Ref. \cite{mirbt} as 1\,Jm$^{-2}$.
The resulting structure of the interfaces inside the superlattice
is shown in Fig. \ref{f2}. From the simulations, the modulation
period of 300\,\AA{} follows, which is in a reasonable agreement
with the observed value $L_{exp}=(267\pm15)$\,\AA{}. It should be
noted that the simulated modulation period is affected by the size
of the simulated region, since there can be only an integer number
of the waves in the simulated system of a given size. To eliminate
the influence of the system size we have simulated the growth of
several systems of sizes 150, 225, 300 and 400\,nm. For various
system sizes, the modulation periods were always obtained in the
interval $(300\pm20)$\,\AA{} depending on the particular system
size. The modulation amplitude is not affected by the system size.

During the growth of first layers in the stack the modulation
amplitude grows exponentially as predicted by the linearized
theory \cite{shilkrot,huang}. In the further growth stage however,
the rate of the growth of the modulation amplitude decreases (see
Fig. \ref{f3}, where the experimentally obtained values of the
first Fourier coefficient $\varepsilon_1$ of $\varepsilon$ are compared
with the simulation results).

The simulations show a good agreement with the experimental
results in spite of the simplified one-dimensional model of the
surface used. Transmission electron microscopy (TEM) on similar
samples \cite{ahrenkiel} revealed that the modulation is nearly
one-dimensional indeed, resulting in a quasiperiodic sequence of
quantum wires; this fact explains why the one-dimensional model is
sufficient for the simulation of the modulation kinetics. The TEM
observations also demonstrated that if the average lattice
parameter of the (relaxed) multilayer is larger than that of the
InP buffer underneath (the actual multilayer structure is
laterally compressed), the modulation direction is close to [100];
if the multilayer is laterally deformed in tension, the modulation
direction is close to the crystallographic directions [310] and
[130]. Of course, the one-dimensional model used here cannot
predict the modulation direction. We ascribe the dependence of the
modulation direction on the deformation sign to the anisotropic
surface tension and anisotropy in elastic constants
\cite{sekerka}. The degree of anisotropy of the surface tension is
also affected by the actual strain in the layer \cite{kaganer} and
this fact could therefore also explain different modulation
directions in the case of a tensile and compressive deformation of
the multilayer.

The continuum simulation also allows for the formation of
non-physical layers the thickness of which are fractional numbers
of monolayers. However, our results based on a continuum
approximation are qualitatively similar to the those obtained
using an atomistic model and a monolayer step corrugation
\cite{bai}.

The observed and predicted modulation periods roughly correspond
to the period given by the linearized theory\cite{shilkrot,huang},
which prediction is $L\approx 200$\,\AA{}.
According to the Ref. \cite{shitara} the surface diffusion of In
is about 50 times faster than Al, although an exact value of the
surface diffusivity of In is not known. In \cite{kasu} the surface
diffusion constant of Al at 530$^\circ$C was found to be
$1.5\times10^{-7}$\,cm$^2$s$^{-1}$. The deposition flux of As
atoms is higher than the flux of In and Al atoms at usual MBE
conditions \cite{herman}, therefore only diffusivities of Al and
In atoms play role.

On the other hand, our simulations have shown that the values of
the diffusion constants have nearly no influence on the modulation
amplitude, since the diffusion process is sufficiently quick and
the growing surface is nearly in an equilibrium state. The
diffusion rate however affects the modulation \emph{period}. In
the case of very slow diffusion (of the order of
$10^{-10}$\,cm$^2$s$^{-1}$ for Al), the growth of the larger ripples at
the expense of smaller ones (the Ostwald ripening)
does not take place and the modulation remains constant during the
growth of the whole superlattice stack. If the diffusion is very
fast (of the order of $10^{-7}$\,cm$^2$s$^{-1}$ for Al), the Ostwald
ripening takes place during the growth of the first layer already,
which leads to the creation of a smaller amount of larger, more
distant dots, separated by larger flat areas of a thin wetting
layer. The nucleation of the ripples on the subsequent interfaces
is affected by the local distribution of lateral strains
originated from the large buried ripples. Due to the elastic
anisotropy, this distribution gives rise to local minima of the
chemical potential at the rims of the buried ripples (two local
minima for each ripple) so that the number of the ripples is
duplicated. After the deposition of several periods, the ripples
cover the whole interface again and the flat areas between the
ripples disappear. The resulting modulation period is approaching
the period obtained for a slow diffusion again.

In our simulations, we have achieved a good correspondence of both
the modulation period and the time dependence of the modulation
amplitude for any value of the diffusion constant of Al between
$10^{-10}$ and $10^{-7}$\,cm$^2$s$^{-1}$.

The resulting interface morphology is substantially affected by
the wetting-effect, i.e., by a non-linear dependence of the volume
density of the elastic energy on the layer thickness. We have
approximated this dependence by Eq. (\ref{wett}). The best
correspondence of the measured and simulated modulation amplitudes
was obtained for the values $E_W=0.15\times E_S$ and
$h_W=0.6\times h_{\rm ml}$. We have also estimated these values by means
of an atomistic simulation of the elastic energy density using the
valence-field force method and the Keating model \cite{keating}.
In these simulations we have neglected the surface relaxation and
reconstruction and we have obtained the dependence of the density
of the elastic energy on the thickness of a layer with a flat
(001) surface. From the fit of this dependence with exponential
formula in Eq. (\ref{wett}) we have obtained $E_W=0.10\times E_S$
and $h_W=0.8\times h_{\rm ml}$, which very well corresponds to the values
above.

The parameters $E_W$ and $h_W$ affect the modulation amplitude and
they have no influence on the modulation period. In the first
stage of the multilayer growth the modulation amplitude rapidly
increases; this increase is slowed down after the growth of about
10 superlattice periods. The parameter $h_W$ affects mainly the
rate of the initial amplitude growth; this rate increase with
decreasing $h_W$. The parameter $E_W$ determines the slowing-down
process: for larger values of $E_W$ the slowdown of the amplitude
growth is observed earlier than for smaller $E_W$.

In conclusion, we have simulated the multilayer growth using a
non-linear continuum model. The simulation results agree very well
with experimental data obtained by x-ray scattering. From the
simulations performed for various values of material parameters we
have found that the wetting effect (the non-linear dependence of
the elastic energy density on the layer thickness) has a crucial
influence on the resulting interface morphology; from the fit of
the experimental data with the simulations we have determined the
parameters of this non-linear dependence and we have compared
these values with atomistic simulations.

\begin{acknowledgments}
The authors are grateful to S.~C.~Moss (University of Houston), A.~Mascarenhas,
A.~G.~Norman (NREL Golden, USA), and J.~L.~Reno (Sandia National Laboratory)
for providing the samples,
B.~Krause and T.~H.~Metzger (ESRF Grenoble) for the assistance
at synchrotron x-ray measurements, and P.~W.~Voorhees (Northwestern University, USA) for helpful discussion.
One of us (O.C.) thanks the
University of Houston for the financial support during his stay in
Houston. The work was supported by the project MSM 0021622410 of
the Ministry of Education of the Czech Republic, and by the NSF through grant DMR-0406323.
K.E.B. is also supported by T$_c$SUH.
\end{acknowledgments}

\bibliography{caha}

\end{document}